\documentclass[aps, prd, showpacs, superscriptaddress, nofootinbib, onecolumn]{revtex4}
\usepackage{amssymb,amsmath,color,microtype}
\usepackage{graphicx}

\usepackage{amssymb,amsmath,amsfonts,amsbsy,graphicx,rotating}

\def\be{\begin{equation}}
\def\ee{\end{equation}}
\def\bea{\begin{eqnarray}}
\def\eea{\end{eqnarray}}
\newcommand{\mpl}{m_{\rm Pl}}

\newcommand{\calP}{{\cal P}}
\newcommand{\calR}{{\cal R}}

\usepackage{color}
\newcommand{\tc}{\textcolor{black}}

\begin{document}

\begin{preprint}
 %\qquad \qquad \qquad \qquad\qquad\qquad\qquad \qquad \qquad \qquad
 APCTP-Pre2014-016
\end{preprint}

\title{On the possibility of blue tensor spectrum within single field inflation}

\author{Yi-Fu Cai}
\email{yifucai@ustc.edu.cn}
\affiliation{CAS Key Laboratory for Research in Galaxies and Cosmology, Department of Astronomy,
University of Science and Technology of China, Chinese Academy of Sciences, Hefei, Anhui
230026, China}
\affiliation{Department of Physics, McGill University, Montr\'eal, Quebec H3A 2T8, Canada}

\author{Jinn-Ouk Gong}
\email{jinn-ouk.gong@apctp.org}
\affiliation{Asia Pacific Center for Theoretical Physics, Pohang 790-784, Korea}
\affiliation{Department of Physics, Postech, Pohang 790-784, Korea}

\author{Shi Pi}
\email{spi@apctp.org}
\affiliation{Asia Pacific Center for Theoretical Physics, Pohang 790-784, Korea}

\author{Emmanuel N. Saridakis}
\email{Emmanuel\_Saridakis@baylor.edu}
\affiliation{Physics Division, National Technical University of Athens, 15780 Zografou
Campus,
Athens, Greece}
\affiliation{Instituto de F\'{\i}sica, Pontificia Universidad de Cat\'olica de
Valpara\'{\i}so,
Casilla 4950, Valpara\'{\i}so, Chile}

\author{Shang-Yu Wu}
\email{loganwu@gmail.com}
\affiliation{Department of Electrophysics, National Center for Theoretical Science,
National Chiao
Tung University, Hsinchu 300, Taiwan}
\affiliation{Shing-Tung Yau Center, National Chiao Tung University, Hsinchu 300, Taiwan}

\pacs{98.80.Cq}

\begin{abstract}
We present a series of theoretical constraints on the potentially viable inflation models
that might yield a blue spectrum for primordial tensor perturbations. By performing a
detailed dynamical analysis we show that, while there exists such possibility, the
corresponding phase space is strongly bounded. Our result implies that, in order to
achieve a blue tilt for inflationary tensor perturbations, one may either construct a
non-canonical inflation model delicately, or study the generation of primordial tensor
modes beyond the standard scenario of single slow-roll field.
\end{abstract}

\maketitle

\section{Introduction}

In recent years, the measurements of the cosmic microwave background (CMB) temperature
anisotropies verified a nearly scale-invariant power spectrum of the primordial curvature
perturbation to high precision \cite{planck}. This observational fact is highly consistent
with the predictions from the perturbation theory of inflationary cosmology
\cite{infpert}. Therefore, inflation, which originally appeared in early 80's
\cite{inflation} (see also \cite{earlyinf}), has become the most prevailing paradigm of
describing the very early universe. Furthermore, inflationary cosmology also predicts
a nearly scale-invariant power spectrum of the primordial gravitational waves, of which
the magnitude is relatively smaller than that of the primordial curvature perturbation
\cite{Starobinsky:1979ty}. {\tc{
If these primordial tensor fluctuations exist, they could lead to the $B$-mode
polarization signals
in the CMB \cite{cmb_b-mode} and hence are expected to be observed in cosmological
surveys.
}}

{\tc{
So far there is no observational evidence that indicates existence of the primordial
tensor fluctuations~\cite{Ade:2015tva, dust_b-mode, Adam:2014bub}. However, as was pointed
out in~\cite{Brandenberger:2014faa}, a suppression of power in the $B$-mode angular power
spectrum at large scales might exist, which implies that a spectrum of primordial
gravitational waves could have a blue tilt. Thus, from the perspective of theoretical
interpretations, it is interesting to investigate whether a power spectrum of
primordial gravitational waves with a blue tilt can be achieved in the framework of
inflationary cosmology.
}}

This question has already drawn the attention of cosmologists in the literature, and a
couple of different mechanisms were put forward, namely the beyond-slow-roll inflation
\cite{beyondsr}, the matter-bounce inflation \cite{Xia:2014tda}, inflation with
non-Bunch-Davis vacuum \cite{Ashoorioon:2014nta}, the non-commutative field inflation
\cite{noncomm_bluetilt}, the variable gravity quintessential inflation
\cite{Hossain:2014xha}, the string gas cosmology \cite{stringgas_bluetilt}, or the Hawking
radiation during inflation~\cite{Mohanty:2014kwa}. Therefore, a careful characterization
of the power spectrum of the primordial $B$-mode polarization is very important to falsify
the paradigms of very early universe (see \cite{Brandenberger:2011eq} for the
characterization of the primordial gravitational waves within various very early universe
models).

In the present work we make a remark on the potential challenge of regular inflation
models to generate a blue tilt for the primordial gravitational waves. We restrict
ourselves within the standard general relativity and present a potential resolution to
this challenge by proposing to extend the parameter space of inflation models by including
non-canonical operators. In particular, we phenomenologically consider a class of
inflation models with the Horndeski operator being involved. Such models were considered
in inflationary cosmology for the purpose of circumventing the paradigm of Higgs inflation
\cite{CervantesCota:1995tz}, and are dubbed as ``G-inflation" \cite{G-inf} (see for
example~\cite{generalG-inf} for generalized analyses and see \cite{Ohashi:2012wf}
for a counter-claim from the stability viewpoint). In our construction, differing from
the application of the Galilean symmetry, inflation is driven by a scalar field with a
Horndeski operator which could be achieved either by the kinetic term or the potential
energy. We investigate the dynamics of this cosmological system by performing a detailed
phase space analysis. We find that in general the generation of a blue tilt of the
primordial gravitational waves in a viable inflation model is difficult since the expected
trajectories are not stable in the phase space. However, a short period of
super-inflationary phase might be possible and thus would circumvent the
above theoretical challenges.

The article is organized as follows. In Section \ref{Sec:general}, we briefly review the
standard picture of predictions made by regular inflation models on the primordial
curvature and tensor perturbations. We can see that it is forbidden to produce a power
spectrum of the primordial tensor modes with a blue tilt in a wide class of inflation
models. Then, in Section \ref{Sec:model} we present a class of extended inflation models
by including a parameterized Horndeski operator. By selecting several typical
inflation potentials, we perform the dynamical analyses in details and derive their
attractor solutions. We show that the inflationary trajectories with a blue tilt do not
correspond to these stable solutions. We conclude with a discussion in Section
\ref{Sec:conclusions}. Throughout the article we take the sign of the metric $(+,-,-,-)$
and define the reduced Planck mass by $\mpl = 1/\sqrt{8\pi G}$.

\section{General discussions}
\label{Sec:general}

In the paradigm of inflationary cosmology, both the primordial curvature perturbation and
gravitational waves are originated from quantum fluctuations in a nearly exponentially
expanding universe at high energy scales. During this expansion, the physical wavelengths
of the metric fluctuations are stretched out of the Hubble radius and then form power
spectra as observed in the CMB. Such a convenient causal mechanism of generating
primordial perturbations can determine their power spectra by a series of simple
relations.

Particularly, for a general inflation model with a k-essence Lagrangian~\cite{k-essence},
the power spectrum of the curvature perturbation $\calR$ is determined by four parameters,
namely the Hubble rate $H$, the spectral index $n_\calR$, the slow-roll parameter
$\epsilon$, and the sound speed parameter $c_s$, through the following relation
\begin{equation}\label{PS}
 \calP_\calR = \frac{\xi_H}{8\pi^2\epsilon c_s} \left( \frac{k}{k_0} \right)^{n_\calR-1}
~,
\end{equation}
with
\begin{equation}\label{xi_H}
 \xi_H \equiv \frac{H_I^2}{\mpl^2}~,
\end{equation}
where $k_0$ is the pivot scale. The subscript $I$ denotes that the value of the Hubble
parameter is taken during the inflationary stage. The slow roll parameter $\epsilon$ is
defined by
\begin{equation}\label{epsilon}
 \epsilon \equiv -\frac{\dot{H}}{H^2}~,
\end{equation}
where dots denote derivatives with respect to $t$, and thus it is determined by the
background dynamics of inflation. The sound speed parameter $c_s$ characterizes the
propagation of primordial scalar fluctuations. Theoretically, its value is constrained
between $0$ and $1$ so that the model is free from the gradient instability and
super-luminal propagation (see however \cite{Babichev:2007dw} for a different viewpoint on
super-luminal propagation of a k-essence field). Moreover, the recent non-detection of the
primordial non-Gaussianity by the Planck data~\cite{Ade:2013ydc} implies that $c_s$ cannot
be too small. The spectral index $n_\calR$ can be derived straightforwardly from its
definition through
\begin{equation}\label{nS}
 n_\calR-1 \equiv \frac{{\rm d}\log \calP_\calR}{{\rm d} \log k} = -4\epsilon+2\eta -s ~,
\end{equation}
where we have introduced two more slow-roll parameters, namely
\begin{equation}\label{eta&s}
 \eta \equiv \epsilon - \frac{\dot{\epsilon}}{2H\epsilon}~, \quad s\equiv \frac{\dot
c_s}{Hc_s}~.
\end{equation}
According to the current CMB observations, the spectral index $n_\calR$ takes a value
which is slightly less than unity and hence the power spectrum of the primordial
curvature perturbation is red-tilted.

For the primordial tensor fluctuations, the associated relations are even simpler if the
gravity theory is still general relativity. The corresponding power spectrum takes the
form of
\begin{equation}\label{PT}
 \calP_T =\frac{2}{\pi^2} \xi_H \left(\frac{k}{k_0}\right)^{n_T}~,
\end{equation}
and therefore it is easy to see that the amplitude of the primordial tensor fluctuations
only depends on the inflationary Hubble parameter and the corresponding spectral index
$n_T$, where the latter, by definition, is given by
\begin{equation}\label{nT}
 n_T \equiv \frac{{\rm d}\log \calP_T}{{\rm d} \log k} = -2\epsilon~.
\end{equation}
Expressions \eqref{PT} and \eqref{nT} are generic to any single field inflation model
minimally coupled to gravity. Since the Hubble rate $H$ is monotonically decreasing in
regular inflation models ($\dot{H}<0$) it is implied that $\epsilon>0$, and hence one can
conclude that the spectral index of the primordial tensor fluctuations $n_T$ in these
models is always negative. Therefore it is red-tilted too, as it is the case of the
curvature perturbation.

Although the present observations cannot determine the tilt of $n_T$, we shall notice that
if one expects a slightly blue power spectrum for the inflationary tensor fluctuations,
$\epsilon$ has to be efficiently negative. This phenomenon implies a violation of
weak/null energy condition during inflation. In the literature there have been some
proposals to give rise to the corresponding energy condition violation in very early
universe, namely super-inflation by the nonlocal gravity approach \cite{Biswas:2013dry},
super-inflation in loop quantum cosmology \cite{superinf_lqg},
inflation in (super-)renormalizable gravity \cite{Briscese:2012ys}, as well as from a
general viewpoint of effective field approach \cite{eft}.

However, it is not trivial to achieve an inflationary model that can realize $\epsilon>0$
in a stable way, without any pathologies, in the framework of Einstein gravity\footnote{It
is known that any single field described by a k-essence type Lagrangian cannot break the
null energy condition without pathologies: see e.g. the appendix of~\cite{Xia:2007km} as
well as a comprehensive review \cite{Cai:2009zp}.}. In particular, one ought to be aware
of the following theoretical constraints:
\begin{itemize}\label{Constraints}
\item
First of all, the model must be stable against any ghost mode, in order for the
perturbation theory describing the primordial perturbations generated from vacuum
fluctuations to be reliable.
\item
The curvature perturbation must be free of the gradient instability, or at least
experience this instability within a very short period. In this regard, there is no
harmful growth of the primordial perturbations that might violate the current
observational constraints.
\item
The spacetime symmetry of the universe should recover the Lorentz symmetry. This indicates
that the theory of matter fields has to recover the canonical version, with all
higher-order operators being suppressed at low energy scales.
\item
After inflation, the universe needs to gracefully exit to the regular thermal expanding
phase smoothly. Hence it is implied that the weak energy condition has to be recovered at
late times of the inflationary stage or after.
\end{itemize}
Keeping these theoretical requirements in mind, it is interesting to look for a viable
inflation model that generates a power spectrum of the primordial tensor fluctuations with
a blue tilt and is consistent with latest cosmological observations. This is exactly the
goal of the present work.

\section{Inflation with a Horndeski operator}
\label{Sec:model}

The violation of the weak energy condition in a stable way, is of theoretical interest in
various models of very early universe physics. One plausible mechanism of achieving such
a scenario is to make use of a ghost condensate field, in which the kinetic term for the
inflaton takes a non-vanishing expectation value in the infrared regime
\cite{ghostcondensate}. However, this type of models often suffers from a gradient
instability, when the universe exits from the inflationary phase to the normal thermal
expansion. Another approach to realize the weak energy condition violation is to make use
of a Galileon-type field (also dubbed as the Horndeski field) \cite{galileon}. The key
feature of this type of field is that it contains higher-order derivative terms
in the Lagrangian, while the equations of motion remain second order and thus do not
necessarily lead to the appearance of ghost modes. These important features have led to
many recent studies of Galileon models which yield a period of inflationary phase at early
times of the universe~\cite{G-inf, generalG-inf, Ohashi:2012wf, Kamada:2013bia}.

In this section, we focus on a class of inflation model with a Horndeski operator. We
phenomenologically consider a dimensionless scalar field $\phi$ with a Lagrangian of the
type
\begin{equation} \label{L_KGB}
 {\cal L} =\frac{\mpl^2}{2}R+ K(\phi, X) + G(\phi, X)\Box\phi~,
\end{equation}
in which $K$ is a k-essence type operator and $G$ is a Horndeski operator. They both are
functions of $\phi$ and the kinetic term
\begin{equation}
 X \equiv \frac{1}{2} g^{\mu\nu} \partial_\mu\phi \partial_\nu\phi  ~,
\end{equation}
and $\Box \equiv g^{\mu\nu}\nabla_\mu\nabla_\nu$ is the standard d'Alembertian operator.
This type of Lagrangian, with specifically chosen forms of $K$ and $G$, was adopted to
drive the late-time acceleration of the universe in \cite{Deffayet:2010qz}, and its
dynamical analysis was carried out in \cite{Leon:2012mt}. Additionally, in
\cite{Cai:2012va} it was found that if one combines the ghost condensate and Horndeski
operators he can obtain a healthy bouncing cosmology, with a smooth transition from a
contracting universe to standard expanding radiation and matter dominated phases (see also
\cite{H-bouncing} for extended studies).

To be specific, we choose the following {\em minimal} ansatz:
\begin{align}
 \label{KG_1}
 K(\phi, X) &= \mpl^2 X - V(\phi) ~,
 \\
 \label{KG_2}
 G(\phi, X) &= \mpl^2 \gamma(\phi) \left( \frac{2X}{\mpl^2} \right)^p ~,
\end{align}
where $\gamma(\phi)$ is a dimensionless function of the inflaton field and $p$ is a
coefficient as a free model parameter. Note that the expression of $K$ in \eqref{KG_1}
corresponds to the canonical Lagrangian for the inflaton field. Moreover, the Horndeski
operator $G$ is expected to stabilize the propagation of the curvature perturbation when
the inflationary stage with $\epsilon<0$ occurs. Its effect is automatically suppressed at
low energy scales if we choose $p$ to be positive definite or $\gamma(\phi)$ to decay
rapidly. Consequently, the model under consideration could partly satisfy the theoretical
limits pointed out at the end of the previous section.

\subsection{Background equations of motion}

Varying the Lagrangian with respect to the metric, and focusing on a flat
Friedmann-Robertson-Walker (FRW) geometry of the form $ ds^2= dt^2 - a^2(t)\,\delta_{ij}
dx^i dx^j $, with $a(t)$ the scale factor, leads to the Friedmann equations
\begin{equation}
\label{FRW_eqs0}
 H^2 = \frac{\rho_\phi}{3\mpl^2}~, \quad \dot{H} = -\frac{\rho_\phi + P_\phi}{2\mpl^2}~,
\end{equation}
where the energy density and the pressure of the scalar field respectively are written as
\begin{align}
 \rho_\phi &= \frac{1}{2} \mpl^2\dot\phi^2 \Big[ 1 + \frac{12p\gamma H\dot\phi}{\mpl^2}
 \big( \frac{\dot\phi^2}{\mpl^2} \big)^{p-1} -2\gamma_\phi \big( \frac{\dot\phi^2}{\mpl^2}
\big)^{p}
 \Big] +V(\phi) ~,
\\
 P_\phi &= \frac{1}{2} \mpl^2\dot\phi^2 \Big[ 1 -\frac{4p\gamma\ddot\phi}{\mpl^2} \big(
\frac{\dot\phi^2}{\mpl^2} \big)^{p-1} -2\gamma_\phi \big( \frac{\dot\phi^2}{\mpl^2}
\big)^{p} \Big] -V(\phi) ~.
\end{align}
Moreover, the equation of motion for the scalar field can be derived as
\begin{equation}
 \label{scalarfieldeq}
 {\cal P} \ddot\phi +{\cal D}\dot\phi +V_\phi = 0~,
\end{equation}
where we have introduced
\begin{align}
\label{P_term1}
 {\cal P} &= \mpl^2 \left[ 1 + \frac{12p^2\gamma H\dot\phi}{\mpl^2} \left(
\frac{\dot\phi^2}{\mpl^2} \right)^{p-1} -2(1+p)\gamma_\phi \left(
\frac{\dot\phi^2}{\mpl^2}
\right)^{p}
 +6p^2\gamma^2 \big( \frac{\dot\phi^2}{M_p^2} \big)^{2p} \right]~,
\\
\label{D_term1}
 {\cal D} &= 3H\mpl^2 \left\{ 1 + \frac{6\gamma H\dot\phi}{\mpl^2} \left(
\frac{\dot\phi^2}{\mpl^2} \right)^{p-1}
  +\left[ 2(p-1)\gamma_\phi - p\gamma\frac{\dot\phi}{H} -
\gamma_{\phi\phi}\frac{\dot\phi}{3H} \right]
\left( \frac{\dot\phi^2}{\mpl^2} \right)^{p}
  + \left( 2\gamma\gamma_\phi \frac{\dot\phi}{H} - 6p^2\gamma^2 \right) \left(
\frac{\dot\phi^2}{\mpl^2}
 \right)^{2p} \right\}~.
\end{align}
Note that, the positivity of the coefficient ${\cal P}$ can be applied to examine whether
the model suffers from a ghost or not. On the other hand, the coefficient ${\cal D}$ is an
effective friction term for the inflaton field. It is easy to check that the regular
Klein-Gordon equation in a FRW background can be recovered if one takes $\gamma=0$.
Finally, for completeness, in Appendix \ref{AppA} we provide the cosmological equations
for general $K(\phi, X)$ and $G(\phi, X)$.

In order to analyze the background dynamics it proves convenient to introduce various
rolling parameters as:
\begin{equation}\label{slowroll1}
 \epsilon_\phi \equiv \frac{\dot\phi^2}{2H^2} ~,~~ \eta_\phi \equiv
-\frac{\ddot\phi}{H\dot\phi} ~,
~~
 \xi_\gamma \equiv \frac{\dot{\gamma}}{H\gamma} ~,~~ \eta_\gamma \equiv
\frac{\dot{\xi_\gamma}}{H\xi_\gamma} ~,
\end{equation}
which are all dimensionless. Note that the first two parameters $\epsilon_\phi$ and
$\eta_\phi$ are mainly associated directly with the dynamics of the inflaton field. In
traditional inflation models with a canonical kinetic term, they coincide with the regular
slow-roll parameters $\epsilon$ and $\eta$ as provided in \eqref{epsilon} and
\eqref{eta&s}. The last two parameters of \eqref{slowroll1} $\xi_\gamma$ and $\eta_\gamma$
respectively describe the first and second order variation of the coefficient $\gamma$
within each Hubble time. In summary, using these parameters one can rewrite the energy
density and the pressure of the scalar field as
\begin{equation}\label{rhoP_phi}
\begin{split}
 \rho_\phi & = V(\phi) + \mpl^2H^2 \left[ \epsilon_\phi +\gamma(6p-\xi_\gamma)
(2\epsilon_\phi)^{
p+1/2}\xi_H^p \right] ~,
 \\
 P_\phi & = -V(\phi) + \mpl^2H^2 \left[ \epsilon_\phi + \gamma(2p\eta_\phi-\xi_\gamma)
(2\epsilon_\phi)^{p+1/2}\xi_H^p \right] ~,
\end{split}
\end{equation}
where $\xi_H$ has been introduced in \eqref{xi_H}.

Recalling that the definition of the background slow-roll parameter $\epsilon$
\eqref{epsilon}, and inserting  \eqref{rhoP_phi} into the Friedmann equations
\eqref{FRW_eqs0}, we can extract the relation between $\epsilon$ and $\epsilon_\phi $ as
\begin{equation}\label{epsilon_general}
 \epsilon = \epsilon_\phi \left[ 1 +2\gamma(3p-\xi_\gamma+p\eta_\phi)
(2\epsilon_\phi)^{p-1/2} \xi_
H^p \right] ~.
\end{equation}
Note that when $\gamma$ vanishes $\epsilon$ is equal to $\epsilon_\phi$ and hence we
reduce to the canonical case where $\epsilon>0$. However, in the general case the dynamics
of $\gamma$ can realize $\epsilon<0$, despite the fact that $\epsilon_\phi>0$. Hence, the
above model can indeed give rise to a blue tilt, as discussed in the previous section. In
the rest of this work we study such a possibility.

\subsection{Cosmological perturbations and ghost avoidance}

A general feature of cosmological scenarios which involve higher-order derivatives is that
they can exhibit ghost instabilities at perturbation level, which can be treated in the
context of an appropriate field redefinition \cite{Gong:2014rna}. Hence, before using
\eqref{L_KGB} for the description of the inflationary phase, one needs to perform a detail
perturbation analysis and extract the necessary conditions for the avoidance of ghosts and
gradient instabilities. Following \cite{noghost} and applying the ansatzes of \eqref{KG_1}
and \eqref{KG_2}, we deduce that in a FRW background the condition for ghost absence
writes as
\begin{equation}
\label{ghostcondition}
 Q_{s} \equiv \frac{w_1 \left( 4w_1w_3+9w_2^{2} \right)}{3w_2^2} \geq 0 ~,
\end{equation}
where
\begin{equation}\label{w_123}
\begin{split}
 w_1 & = w_4 = \mpl^2 \, ,
 \\
 w_2 & = 2 G_XX\dot{\phi} + 2\mpl^2H = 2\mpl^2 \Big[ H +p\gamma\dot{\phi} \big(
\frac{\dot{\phi}^2}{
\mpl^2} \big)^p \Big] \, ,
 \\
 w_3 & = -9\mpl^2H^2 + 3X \left( K_X + 2X K_{XX} \right)
 + 6X \left( G_\phi + X G_{\phi X} - 6 H\dot{\phi}G_{X} - 3XH\dot{\phi}G_{XX} \right)
 \\
 & = \frac{3}{2}\mpl^2\dot{\phi}^2 \Big[ 1  + (p+1) \big( \frac{\dot{\phi}^2}{\mpl^2}
\big)^p \big(
\gamma_\phi - 6p\gamma\frac{H}{\dot{\phi}} \big) \Big] - 9\mpl^2H^2 ~.
\end{split}
\end{equation}
{\tc{Note that the physical meaning of the $Q_s$ parameter corresponds to the positivity
coefficient of the perturbation variable which appears in Eq. \eqref{z^2_app} in
Appendix \ref{AppB}.
}}

In addition, the condition for the avoidance of gradient instabilities (associated with
the scalar field propagation speed) reads
\begin{equation}
\label{Laplaciancondition}
 c_{s}^{2}\equiv \frac{3 \left( 2w_1^2w_2H-w_2^2w_4+4w_1w_2\dot{w}_1-2w_1^2 \dot{w}_2
\right)}{w_1 \left( 4w_1w_3+9w_2^2 \right)} \geq 0 ~.
\end{equation}
{\tc{
Note that the above two theoretical constraints can impose the bound of the parameters
for inflation models in the literature directly. To be explicit, the condition for ghost
absence \eqref{ghostcondition} requires that $$w_3 \geq -\frac{9w_2^2}{4w_1},$$
and the condition for the avoidance of gradient instabilities \eqref{Laplaciancondition}
implies $$ 0 \leq w_2 \leq 2H w_1, $$ under the assumption of $|\dot{w}_2/w_2H|\ll1$. As
a result, the parameter space of inflation models that attempt to generate a blue
spectrum of primordial gravitational waves could be strongly constrained by the above two
theoretical requirements.
}}
{\tc{
Furthermore, we would like to provide several examples of explicit inflation models, in
order to demonstrate how these theoretical conditions can constrain the validity of
these models. For instance, for a model of canonical inflation with $\gamma=0$, these two
conditions are satisfied automatically, since $w_2=2Hw_1$ exactly, however the
spectrum of
tensor modes is also red tilted; moreover, if we consider an explicit model with $p=1$
and
$\gamma = \gamma_0 e^{\lambda_\gamma \phi}$ and take a positive value for $\dot\phi$,
then the theoretical constrains yield approximately that $\gamma_0\leq0$ and
$2\gamma_0\lambda_\gamma e^{\lambda_\gamma\phi}\dot\phi^2+\mpl^2>0$. For the latter case,
there might exist possible parameter space that allows for a blue tilt for tensor modes,
however, as will be shown in the following subsection, this possibility is unstable under
the
phase space analysis.
}}

We refer to Appendix~\ref{AppB} for a specific instruction of the perturbation analysis
for the inflation model under consideration.

\subsection{Dynamical analysis}

Let us now apply the powerful method of dynamical analysis
\cite{infexit,phasespaceanalysis,phasespaceanalysis1} in order to
investigate the general features of inflation in the scenario at hand. In order to perform
such a stability analysis we first transform the cosmological equations in their
autonomous form $ \bf{X}'=\bf{f(X)}$, where primes denote derivative with respect to $\log
a$, with $\bf{X}$ a vector constituted by suitable auxiliary variables and $\bf{f(X)}$ the
corresponding vector of the autonomous equations. The critical points $\bf{X_c}$ of this
autonomous system are extracted through the condition $\bf{X}'=0$. Their stability is
examined by expansion around them as $\bf{X}=\bf{X_c}+\bf{V}$, with $\bf{V}$ the vector of
the variable perturbations, resulting to the perturbation equations of the form
$\textbf{V}'={\bf{Q}}\cdot \textbf{V}$, with the matrix ${\bf {Q}}$ containing all
the coefficients of these equations. Therefore, the type and properties of a specific
critical point are determined by the eigenvalues of ${\bf {Q}}$: eigenvalues with positive
real parts imply instability, eigenvalues with negative real parts imply stability, while
eigenvalues with real parts of different sign correspond to a saddle point. In this way,
one is able to extract qualitative information for the global dynamics of the examined
scenario, independently of the initial conditions and the specific universe evolution.

We are interesting in analyzing the Friedmann equations (\ref{FRW_eqs0}), along with the
scalar field evolution equation (\ref{scalarfieldeq}). Considering for simplicity the
most important case where $p=1$, we introduce the auxiliary variables
\begin{equation}\label{auxilvar}
 x \equiv \frac{\dot{\phi}}{\sqrt{6}H} ~, \quad y \equiv
\frac{\sqrt{V(\phi)}}{\sqrt{3}\mpl H} ~, \quad z \equiv \frac{\gamma(\phi) H
\dot\phi}{\mpl^2}~.
\end{equation}
In terms of these variables the first Friedmann equation takes the form
\begin{equation}\label{Constraint}
 (1+12 z) x^2+y^2-2\sqrt{6}zx^3\frac{\gamma_\phi}{\gamma}=1 \, ,
\end{equation}
while from the definitions of $x$ and $z$ we acquire  $H^2=\mpl^2 z/\left(\sqrt{6}
x\gamma\right)$.

In order to proceed the analysis we have to consider ansatzes for the potential $V(\phi)$
and the Galileon coupling function $\gamma(\phi)$. As a simple model we analyze the
exponential potential
\begin{equation}
\label{exponentialpot}
 V(\phi)=V_0 e^{\lambda_V\phi},
\end{equation}
which is widely used in the literature~\cite{expV,phasespaceanalysis}, along with an
exponential coupling
function
\begin{equation}\label{exponentialcoupling}
 \gamma(\phi) = \gamma_0 e^{\lambda_\gamma \phi} ~.
\end{equation}
{\tc{
We comment that the specific forms considered above are very representative and can grasp
common features obtained in other cases too. Firstly, the exponential potential is one of
the most representative ansatzes used in cosmology. In particular, this exponential
potential possesses a manifest advantage that it can mimic any arbitrary cosmic evolution
with a constant equation of state. Additionally, it is well known that an exponential
potential can yield an attractor solution. Secondly, from the discussion performed below
Eq. \eqref{Laplaciancondition}, one can see that one important theoretical bound for
$\gamma$ is its positivity. This can be exactly
implemented by taking the form of $\gamma$ to be exponential, and accordingly, its
positivity is only determined by the $\gamma_0$ parameter without any sign change. As will
be shown in the following dynamical analysis, the evolution of the inflaton field under
this specific form can allow for fixed points in the trajectories of the inflaton field.
}}

In this case, \eqref{FRW_eqs0} and \eqref{scalarfieldeq} are transformed to the autonomous
form:
\begin{align}
\label{equationx}
 x' & = \left\{ 2\left[1 +4z \left( 3 +9x^2z -\sqrt{6}x \lambda_\gamma \right) \right]
\left[ -x^2
+2zx^2 \left( \sqrt{6}x\lambda_\gamma-6 \right) -y^2 \right] \right\}^{-1}
 \nonumber\\
 & \quad \times \left\{ 6 x y^2 (1 + 6 z) +\sqrt{6}\lambda_V y^4 +\sqrt{6}\lambda_V x^2
y^2 (1 + 6
z) + 24\sqrt{6} \lambda_\gamma x^4 z \left[ 1 + 3 (3 + y^2) z \right]
\right.
 \nonumber\\
 & \qquad \ \ + 24\sqrt{6} \lambda_\gamma x^6 z^2 \left( 6 + 54 z + \lambda_\gamma^2
\right)
 - 12x^3z \left[3 +18z +y^2 \left( 3+18z +\lambda_\gamma \lambda_V +\lambda_\gamma^2
\right) \right]
 \nonumber\\
 & \qquad \ \ - 12x^5z \left[3 +\lambda_\gamma^2 +6z(9 + 36 z + 5 \lambda_\gamma^2)
\right]  -864\lambda_\gamma^2 x^7 z^3 \Big\} \, ,
 \\
\label{equationy}
 y' & = \mpl x y \left\{ 2\left[1 +4z \left( 3 +9x^2z -\sqrt{6}x \lambda_\gamma \right)
\right] \left[ -x^2 +2zx^2 \left( \sqrt{6}x\lambda_\gamma-6 \right) -y^2 \right]
\right\}^{-1}
 \nonumber\\
 & \quad \times \Big\{ 432\lambda_V \lambda_\gamma x^5 z^3-\sqrt{6} \lambda_V  y^2 (1 + 18
z) + 36
x^3 z \lambda_\gamma (\lambda_V +12z\lambda_V -6z\lambda_\gamma)
 \nonumber\\
 & \qquad  \ \ - \sqrt{6}x^2 \left\{ \left[ 1+12z \left(2+12z+3zy^2 \right) \right]
\lambda_V - 36\lambda_\gamma z (1 + 8 z) \right\}
 \nonumber\\
 & \qquad   \ \ - 12\sqrt{6} x^4 z^2 \lambda_V \left( 3 + 36 z + 4 \lambda_\gamma^2
\right) - 6x\left[ 1+4z \left( 6 +27z -y^2\lambda_V\lambda_\gamma \right) \right] \Big\}
~,
 \\
\label{equationz}
 z' & = z\left\{ 2x \left[ 1 + 4 z (3 + 9 x^2 z - \sqrt{6} x  \lambda_\gamma ) \right]
\right\}^{-1}
 \nonumber\\
 & \quad \times \Big\{ 6x \left\{ -1 -6z -x \left[ x +36x z^2 \left( 2
-\sqrt{6}\lambda_\gamma x + \lambda_\gamma^2 x^2 \right) \right.\right.
 \nonumber\\
 & \left.\left. \qquad\qquad\ \, + z \left( \sqrt{6} y^2 \lambda_V +18x
-6\sqrt{6}\lambda_\gamma
x^2 - 2\lambda_\gamma^2 x \right) \right] \right\} - \sqrt{6} \lambda_V y^2 \Big\} ~.
\end{align}
Additionally, in terms of the auxiliary variables, the energy density and pressure of the
scalar field can be rewritten as follows:
\begin{align}
 \rho_\phi & = 3\mpl^2H^2 \left\{ y^2 + x^2 \left[ 1-2z \left( \sqrt{6}x\lambda_\gamma-6
\right) \right] \right\} ~,
 \\
 P_\phi & = \frac{3\mpl^2H^2}{ 4z \left( 9x^2z -\sqrt{6}\lambda_\gamma x+3 \right)+1 }
\nonumber\\
 & \quad \times \left\{ 24\lambda_\gamma^2x^4z^2 - 6\sqrt{6}(1+4 z)z\lambda_\gamma x^3  +
\left[
12z(3z+2)+1 \right] x^2 + 2\sqrt{6}y^2xz (2\lambda_\gamma +\lambda_V) - y^2(12z+1)
\right\} ~,
\end{align}
and hence the equation of state of the scalar field takes the form
\begin{equation}
\label{wphiaux}
 w_\phi = \frac{ 24\lambda_\gamma^2 x^4 z^2 -6\sqrt{6} (1+4 z) z \lambda_\gamma x^3  + [12
z(3 z+2)
+1] x^2 -y^2 (12 z+1) + 2\sqrt{6} y^2 x z (2 \lambda_\gamma+\lambda_V) }{ \left\{y^2 + x^2
\left[1 -
 2 z \left( \sqrt{6} x \lambda_\gamma-6 \right)\right]\right\} \left[4z \left(9 z x^2-
\sqrt{6} \lambda_\gamma x+3\right)+1\right] } \, .
\end{equation}
Note that knowing $w_\phi$ we can straightforwardly calculate the  deceleration parameter
as $q\equiv -1 -\dot{H}/H^2 =1/2+3w_\phi/2$. Finally, we can express the two
instability-related
quantities $Q_s$ and $c_s^2$ given in \eqref{ghostcondition} and
\eqref{Laplaciancondition} in
terms of the auxiliary variables as
\begin{align}
\label{ghostcondition2}
 Q_s & = \frac{3 x^2 \left[4z \left(9 z x^2-  \sqrt{6} \lambda_\gamma
x+3\right)+1\right]}{\left(6
z x^2-1\right)^2} ~,
 \\
\label{Laplaciancondition2}
 c_s^2 & = \left\{ {x \left[ 4z \left(9 z x^2-  \sqrt{6} \lambda_\gamma x+3 \right)+1
\right]^2} \right\}^{-1} \left\{ 12 z^2x^3 \left(2 \lambda_\gamma^2+12 z+5 \right)
-24\sqrt{6} z^3 \lambda_\gamma x^4 \right.
 \nonumber\\
 & ~~ \left. - 432 z^4 x^5 -4\sqrt{6} (1+8 z) z \lambda_\gamma x^2 +x + 4xz \left[
3z\left(5-3 y^2\right) +2\right]-2\sqrt{6} y^2 z \lambda_V \right\} ~.
\end{align}
The real and physically meaningful critical points (namely those that correspond to an
expanding universe, i.e. with $H>0$) of the autonomous system \eqref{equationx} -
\eqref{equationz} are obtained by setting the left hand sides of these equations to zero,
and are presented in Table~\ref{Table1} along with their existence conditions. For each of
these critical points we calculate the $3\times3$ matrix ${\bf Q}$ of the perturbation
equations as we described above, and we extract its eigenvalues which are given in
Table~\ref{Table1} too. Hence, we use them in order to deduce the stability properties.
Furthermore, since we have the coordinates of the critical points of the autonomous system
at hand, we can use them to calculate the corresponding $w_{\phi}$ and $q$ from
\eqref{wphiaux}, as well as $Q_{s}$ and $c_{s}^{2}$ from \eqref{ghostcondition2} and
\eqref{Laplaciancondition2} respectively, and we present them in Table~\ref{Table2}.
Finally, using the obvious relation $\epsilon=q+1$, in the last column of
Table~\ref{Table2} we present $\epsilon$ of the corresponding
critical points.
\begin{table}[!]
%\begin{sidewaystable}[!]
\begin{tabular}{|c||c|c|c|c|c|c|}
\hline
 Points & $x_c$ & $y_c$ &  $z_c$ & Exist for & Eigenvalues & Stability
\\
\hline \hline
$A$& $+1$ & $0$ & $0$& always &  3, $\sqrt{6} \lambda_\gamma-6$, $3+\sqrt{\dfrac{3}{2}}
\lambda_V$
   & {\small{unstable for $\lambda_V>-\sqrt{6}$, $\lambda_\gamma>\sqrt{6}$ }}
\\
&&&& & & saddle point otherwise
\\
\hline
$B$& $-1$ & $0$ & $0$& always & 3, $-\sqrt{6} \lambda_\gamma-6$, $3-\sqrt{\frac{3}{2}}
\lambda_V$
   & {\small{unstable for $\lambda_V<\sqrt{6}$, $\lambda_\gamma<-\sqrt{6}$ }}
\\
&&&& & & saddle point otherwise
\\
\hline
&&&& &  & {\small{unstable for $\lambda_\gamma>\sqrt{6}$, $\lambda_V>-\lambda_\gamma$ }}
\\
 $C$& {\small{$\dfrac{ \lambda_\gamma+\sqrt{{\lambda_\gamma}^{2}-6}}{\sqrt{6}}$}} & $0$ &
{\small{
$\dfrac{\alpha^--3}{18}$}} & $\lambda_\gamma^2\geq 6$  &  $C^-$, $C^+$,
$\dfrac{(\lambda_\gamma+\lambda_V) \alpha^+}{2 \lambda g}    $      & saddle point
otherwise
\\
\hline
&&&& &  & {\small{unstable for $\lambda_\gamma<-\sqrt{6}$, $\lambda_V<-\lambda_\gamma$ }}
\\
 $D$& {\small{$\dfrac{\lambda_\gamma-\sqrt{{\lambda_\gamma}^{2}-6}}{\sqrt {6}}$}} & $0$
&{\small{ $\dfrac{\alpha^+-3}{18}$}} & $\lambda_\gamma^2\geq 6$   & $D^-$,
$D^+$,$\dfrac{(\lambda_\gamma+\lambda_V) \alpha^-}{2 \lambda g}$  & saddle point otherwise
  \\
\hline
&&&& & $\lambda_V^2-3$, $-\lambda_V (\lambda_\gamma+\lambda_V)$, & {\small{stable node for
$-\sqrt{
3}<\lambda_V<0$, $\lambda_\gamma<-\lambda_V$ }}
\\
 $E$ & $-\dfrac{ \lambda_{{V}}}{\sqrt {6}}$ &
{\small{$\sqrt{1-\dfrac{{\lambda_{{V}}}^{2}}{6}}$ }} &
 $0$ &{\small{ $0<\lambda_V^2\leq 6$}} & $\dfrac{\lambda_V^2-6}{2}$    &  {\small{stable
node for
$0<\lambda_V<\sqrt{3}$, $\lambda_\gamma>-\lambda_V$ }}
 \\
&&&& &   &saddle point otherwise
\\
\hline
 $F$& $0$ & $1$  & $0$ & $\lambda_V=0$ & $-3$, $-3$, $-3$ & stable node \\
\hline
\end{tabular}
\caption[Table1]{\label{Table1} Real and physically meaningful critical points of the
autonomous
system \eqref{equationx}-\eqref{equationz}, their existence conditions, their
corresponding
eigenvalues and their stability conditions, in the case of exponential potential
\eqref{exponentialpot} and exponential coupling function \eqref{exponentialcoupling}. We
have defined $
\alpha^\pm(\lambda_\gamma) \equiv \lambda_\gamma^2 \pm \lambda_\gamma
\sqrt{\lambda_\gamma^2-6} > 3 $, as well
as $C^{\pm} \equiv 3 \left[\pm\sqrt{2} \lambda_\gamma \left( \alpha^--3 \right)^{-1/2} +
\alpha^+ -
4 \right]/4$ and $D^{\pm} \equiv 3 \left[ \pm \sqrt{2} \lambda_\gamma \left( \alpha^+-3
\right)^{-1/
2} + \alpha ^- -4 \right]/4$. }
%\end{sidewaystable}
\end{table}

\begin{table}[!]
\begin{center}
\begin{tabular}{|c||c|c|c|c|c|}
\hline
 Points   & $w_\phi$ & q & $c_S^2$ & $Q_S$  &$\epsilon$
\\
\hline \hline
$A$   & $1$& $2$& $1$& $3$ &  3
\\
\hline
$B$  & $1$& $2$& $1$& $3$& 3
\\
\hline
 $C$  & $-1 +\dfrac{\alpha^+}{3}$ & $-1+\dfrac{\alpha^+}{2}$ & $0$& arbitrary  &
$\dfrac{\alpha^+}{
2}$
\\
\hline
 $D$   & $-1 +\dfrac{\alpha^-}{3}$& $-1 +\dfrac{\alpha^-}{2}$ & $0$& arbitrary  &
$\dfrac{\alpha^-}{
2}$
\\
\hline
 $E$ & $-1 +\dfrac{\lambda_V^2}{3}$ & $-1 +\dfrac{\lambda_V^2}{2}$ & $1$
&$\dfrac{\lambda_V^2}{2}$&
 $\dfrac{\lambda_V^2}{2}$
\\
\hline
 $F$ & $-1$ & $-1 $ & $1$ &$0$&0 \\
\hline
\end{tabular}
\end{center}
\caption[Table2]{\label{Table2} Real and physically meaningful critical points of the
autonomous system \eqref{equationx} - \eqref{equationz}, and the corresponding values of
the scalar field equation of state $w_\phi$, the deceleration parameter $q$ and the
instability-related parameters $c_s^2$ and $Q_s$, which must be non-negative for a
scenario free of ghosts and gradient instabilities, and the slow-roll parameter
$\epsilon$.}
\end{table}

Let us now discuss the physical behavior of the above dynamical analysis. Since in this
work we investigate inflation realization, first of all we are interested in those
critical points where the expansion of the universe is accelerating, especially those with
$q \approx -1$. Amongst them, we are interested in those points that are saddle or
unstable, which means that if the universe starts from them, i.e. from inflation, the
dynamics will naturally lead the universe away from them, viz. it will offer a natural
exit from inflation~\cite{infexit,infexit2}.
%We mention that this is the opposite of the late-time investigation, where we are
%interested in
%stable critical points, i.e. in points that can attract the universe at late times
%independently of
%the initial conditions.

As we can see from Tables~\ref{Table1} and \ref{Table2}, point $E$ exhibits these
features, and thus it corresponds to the inflationary solution we are looking for. Note
that the physical quantities depend only on the potential exponent $\lambda_V$.
In particular, the smaller $\lambda_V$ is, the closer we are to de Sitter phase. Finally,
note that both $c_s^2$ and $Q_s$ are positive there, which means that this inflationary
solution is free of ghosts and potential instabilities. However, in this solution we
obtain $w_\phi = -1+\lambda_V^2/3$, which is always larger than $-1$, and thus
correspondingly $\epsilon$  is always positive definite along the stably inflationary
trajectory (note that $\epsilon\geq0$ in all the obtained points). Therefore, $n_T$ is
not allowed to be positive in the model under consideration.

\begin{figure}[!]
\begin{center}
\includegraphics[width=9cm]{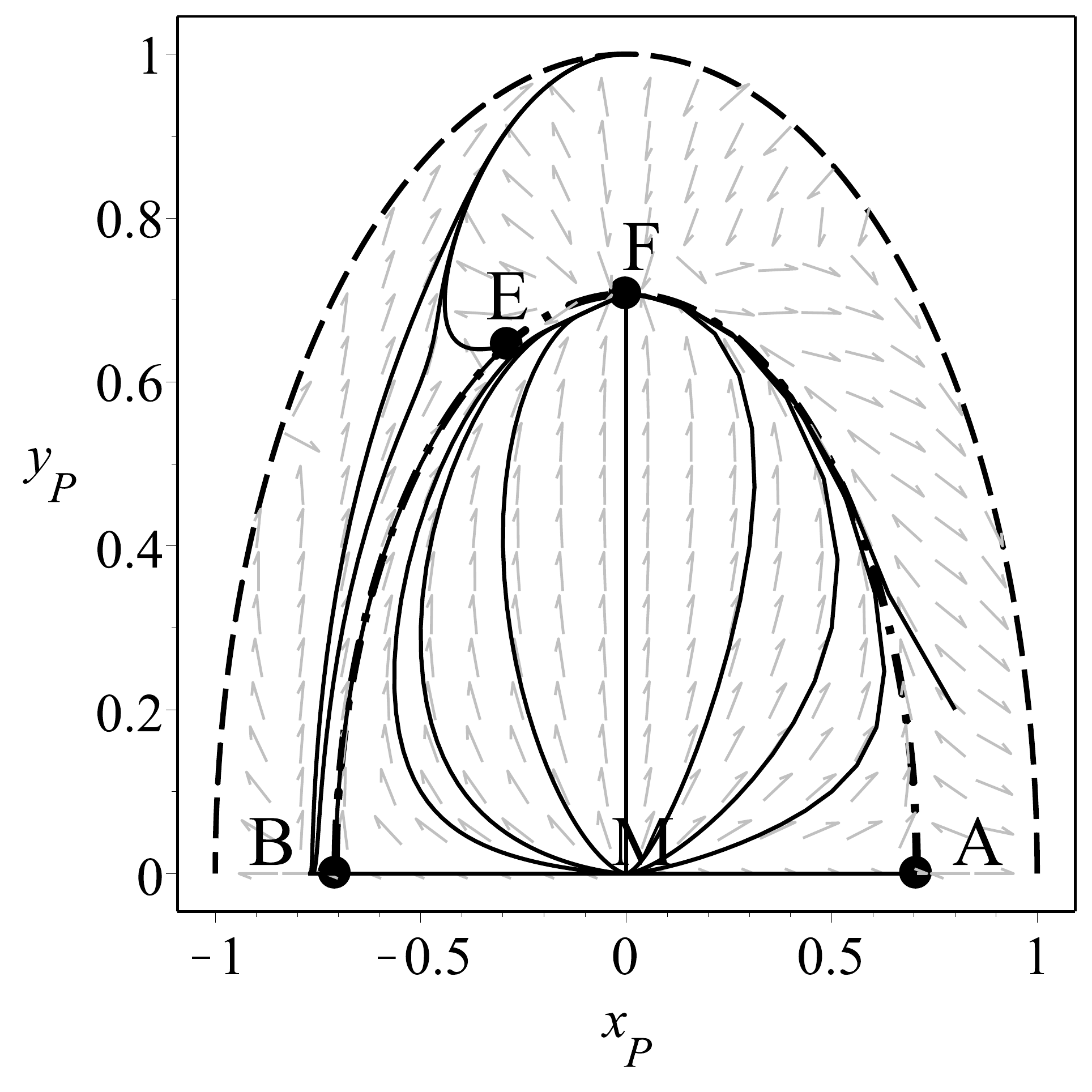}
\end{center}
\caption{Projection on the $x_P$-$y_P$ plane of the phase space behavior of the model
\eqref{KG_1}, with $V(\phi)=V_0 e^{\lambda_V\phi}$ and $\gamma(\phi) = \gamma_0
e^{\lambda_\gamma \phi}$, for $p=1$, $\lambda_V=1$ and $\lambda_\gamma=-2$. The region
inside the inner semi-circle (seen as semi-ellipse in the figure scale), marked by the
thick dashed-dotted line, is the physical part of the phase space. In this projection
point $E$ is saddle, $F$ is an attractor, $A$ and $B$ are unstable, and the origin $M$ is
a saddle. The inflationary realization is described by point $E$. }
\label{Fig1}
\end{figure}

In order to see this behavior more transparently, we evolve the autonomous system
\eqref{equationx} - \eqref{equationz} numerically for the choice $\lambda_V=1$ and
$\lambda_\gamma=2$ and we present the resulting phase space behavior in Figure~\ref{Fig1}.
For convenience, we project the phase space trajectories on the $x_P$-$y_P$ plane of the
Poincar\'e variables $x_P=x/ \sqrt{1+x^2+y^2}$ and $y_P=y/ \sqrt{1+x^2+y^2}$. As we can
observe, the realization of inflation is described by the saddle point $E$, and the
departure of the system from it after a finite time corresponds to the exit from
inflation.

\section{Conclusions}
\label{Sec:conclusions}

In the present article we have studied in detail the theoretical challenge of single
field inflation models to generate a blue tilt for the primordial gravitational waves.
Considering a generalized single field inflation model with a Horndeski operator minimally
coupled to Einstein gravity, we have performed a detailed phase space analysis and have
shown explicitly that the only inflationary solution without any pathologies yields a
positive definite value of the slow-roll parameter $\epsilon$. Therefore, up
to leading order, the spectral index of the primordial tensor perturbations, which takes
the form of $-2\epsilon$ under the consistency relation, is always red tilted.

There might be directions to circumvent the theoretical difficulty pointed out in the
present study, however this would require to extend into more complicated situations. For
instance, one could try to go beyond the single slow-roll field. For instance,
relation \eqref{nT} can be altered by taking into account possible contributions that are
higher-order in slow-roll~\cite{beyondsr}, by including particle production
effects~\cite{Mukohyama:2014gba}, or by considering inflation models driven by some
non-conventional matter such as in elastic inflation~\cite{Gruzinov:2004ty}
and solid inflation~\cite{Endlich:2012pz}. However, the analyses of these possible
complicated scenarios under the theoretical constraints imposed in
Section~\ref{Sec:general}, need to be
performed in detail in future projects, before accepting them as successful candidates for
the description of Nature.

We end the present paper by clarifying our motivation. In this paper we pointed out a
theoretically severe problem. That is, given a possible detection of a blue spectrum for
primordial tensor fluctuations, single field inflation models under our current knowledge
can hardly provide a reasonable interpretation. To demonstrate this theoretical
difficulty, we have performed the dynamical system analysis in detail based on a class of
G-inflation model of which the form is pretty generic. This issue might be circumvented by
more complicated choices of the functions $K$ and $G$, but more problems associated with
other instabilities/inconsistencies would appear, such as observational constraints of
primordial non-gaussianities. We argue that, if the difficulty of realizing a blue tensor
spectrum could eventually become a no-go theorem, then such a possible detection may spoil
the picture of single field inflationary cosmology completely; otherwise, a
more delicate model building is required. This is the goal of our following-up project.

\subsection*{Acknowledgments}

We thank Robert Brandenberger, Genly Leon, Jerome Quintin and Yi Wang for useful
discussions. YFC thanks the Asia Pacific Center for Theoretical Physics for warmest
hospitality during his visit. YFC is supported in part by the ``Thousand Talent Program
for Young Outstanding Scientists'' of China, by the USTC start-up funding under Grant No.
KY2030000049, by the Natural Sciences and Engineering Research Council (NSERC) of Canada
and by the Department of Physics at McGill. JG and SP acknowledge the
Max-Planck-Gesellschaft, the Korea Ministry of Education, Science and Technology,
 Gyeongsangbuk-Do and Pohang City for the support of the Independent Junior Research Group
at the Asia Pacific Center for Theoretical Physics. They are also supported in part by a
Starting Grant through the Basic Science Research Program of the National Research
Foundation of Korea (2013R1A1A1006701). The research of ENS is implemented within the
framework of the Operational Program ``Education and Lifelong Learning'' (Actions
Beneficiary: General Secretariat for Research and Technology), and is co-financed by the
European Social Fund (ESF) and the Greek State. SYW was supported by Ministry of Science
and Technology and the National Center of Theoretical Science in Taiwan.
						
%\newpage

\appendix

\section{Background dynamics}
\label{AppA}

In this Appendix  we provide the general forms of the equations of motion in the model
under consideration. Varying the Lagrangian (\ref{L_KGB}) with respect to the metric, one
can obtain the Friedmann equations that determine the dynamics of the background universe
as
\begin{equation}
 H^2 = \frac{\rho_\phi}{3\mpl^2}~, \quad \dot{H} = -\frac{\rho_\phi+P_\phi}{2\mpl^2}~,
\end{equation}
where in the general case the energy density and pressure write as
\begin{align}
 \rho_\phi &= 2X K_{X} -K +6G_{X}H\dot\phi X -2XG_{\phi} ~,
 \\
 P_{\phi} &= K -2X \left( G_{\phi} +G_{X}\ddot\phi \right) ~,
\end{align}
respectively. In addition, varying (\ref{L_KGB}) with respect to the scalar field yields
the generalized Klein-Gordon equation
\begin{eqnarray}
 && K_{X} \left( \ddot\phi+3H\dot\phi \right) + 2K_{XX}X\ddot\phi +2K_{\phi X}X -K_{\phi}
 - 2\left( G_{\phi}-G_{\phi X}X \right) \left( \ddot\phi+3H\dot\phi \right) -4G_{\phi
X}X\ddot\phi
 \nonumber\\
 &&  \ \ \ \ \ \ \ \ \ \ \ \ \ \ \ \ \ \ \ \ \  -2G_{\phi\phi}X +6G_{X} \left[
(HX)^{\cdot}+3H^2X \right] +6G_{XX}HX\dot{X} = 0 ~,
\end{eqnarray}
which is a second order differential equation and hence it is free of extra degrees of
freedom. Now we can insert specific operators $K$ and $G$, for example \eqref{KG_1} and
\eqref{KG_2}, into the above equations, and derive out straightforwardly the detailed
equations of motion.

\section{Perturbation dynamics}
\label{AppB}

In this Appendix we present the detailed expression of the quadratic action that
characterizes the dynamics of the cosmological perturbations at linear order. For the
curvature perturbation, the quadratic action is given by
\begin{equation}\label{S_2_calR}
 S_2 = \int dt d^3x \frac{a}{2}z^2 \left[ \dot\calR^2 - \frac{c_s^2}{a^2}(\nabla\calR)^2
\right] ~,
\end{equation}
where $z^2$ and $c_s^2$ are given by
\begin{align}
\label{z^2_app}
 z^2 & = \frac{4a^2\mpl^4\dot\phi^2 }{\left( 2\mpl^2H - \dot\phi^3G_{X} \right)^2}
 \left[ K_{X} +\dot\phi^2 K_{XX} +6H\dot\phi G_{X} + \frac{3\dot\phi^4G_{X}^2}{2\mpl^2}
+3H\dot\phi^3G_{XX} -2G_{\phi} -\dot\phi^2G_{\phi X} \right] ~,
\\
\label{c_s^2_app}
 c_s^2 & = \frac{ K_{X} + 4H\dot\phi G_{X} -\dot\phi^4G_{X}^2/\left(2\mpl^2\right)
-2G_{\phi} +\dot\phi^2G_{\phi X} +\left( 2G_{X}+\dot\phi^2G_{XX} \right)\ddot\phi }{ K_{X}
+\dot\phi^2 K_{XX} +6H\dot\phi G_{X} + 3\dot\phi^4G_{X}^2/\left(2\mpl^2\right)
+3H\dot\phi^3G_{XX} -2G_{\phi} -\dot\phi^2G_
{\phi X} } \, ,
\end{align}
{\tc{
where the expression of $z^2$ is related to Eq. \eqref{ghostcondition} via $z^2 = 2a^2 Q_s$.
}}

Under the specific $ K(\phi, X)$ and $ G(\phi, X) $ ansatzes of (\ref{KG_1}) and
(\ref{KG_2}), as well as the slow-roll approximation, the above coefficients can be
significantly simplified to leading order as
\begin{align}
\label{z2_cs2}
 z^2 & \approx 2a^2\mpl^2\epsilon_\phi \frac{1
+12p^2\gamma(2\epsilon_\phi)^{p-1/2}\xi_H^p}{ \left[
1 -p\gamma(2\epsilon_\phi)^{p+1/2}\xi_H^p \right]^2} ~,
\\
 c_s^2 & \approx \frac{1 +8p\gamma(2\epsilon_\phi)^{p-1/2}\xi_H^p}{1
+12p^2\gamma(2\epsilon_\phi)^{
p-1/2}\xi_H^p} ~.
\end{align}
We mention that $z^2$ is required to be positively definite in order for the model to be
free of any ghost mode. During inflation, $c_s^2$ is also required to be positive and
therefore the propagation of the primordial perturbations do not suffer from a gradient
instability.

\end{document}